\documentclass[aps,preprint,groupedaddress]{revtex4}
\usepackage{epsfig,amsbsy,bm}
\newcommand{\eref}[1] {(\ref{#1})}
\newcommand{\Eref}[1] {Eq.~(\ref{#1})}

\newcommand{\isum}%
{\mathop{\hbox{$\displaystyle\sum\kern-13.2pt\int\kern1.5pt$}}}

\begin{document}
\bibliographystyle{apsrev}
\baselineskip = 8mm

\title{Lippmann-Schwinger description of  multiphoton
ionization}

\author{I. A. Ivanov\footnote[1]{Corresponding author:
Igor.Ivanov@.anu.edu.au}\footnote[2]{On leave from the Institute of
Spectroscopy, Russian Academy of Sciences}
 and A. S. Kheifets}
\affiliation
{Research School of Physical Sciences and Engineering,
The Australian National University,
Canberra ACT 0200, Australia}

\date{\today}
\begin{abstract}
We outline  a formalism and develop a computational procedure to treat the
process of multiphoton ionization (MPI) of atomic targets in strong
laser fields. We treat the MPI process nonperturbatively as a decay
phenomenon by solving a coupled set of the integral Lippmann-Schwinger
equations. As  basic building blocks of the theory we use a
complete set of field-free atomic states, discrete and
continuous. This approach should enable us to provide both the total
and differential cross-sections of MPI of atoms with one or two
electrons.  As an illustration, we apply the proposed procedure to a simple
model of MPI from a square well potential and to the hydrogen atom.
\end{abstract}

\maketitle

\section{Introduction}
\label{S1}

In recent years, the process of multiphoton ionization (MPI) of atomic
and molecular species has been a subject of intensive experimental and
theoretical studies (see reviews by \citet{PKK97},
\citet{LMZ98}, \citet{CT04} and \citet{Posthumus04}
and references therein).  Rapid progress in this field has been
largely driven by advancement in high-power short-pulse laser
techniques. The laser intensities which may go beyond
$10^{13}$~Wcm$^{-2}$ make it possible to observe many striking
phenomena such as MPI and above-threshold ionization.

Accurate theoretical description of ionization processes occurring in
laser fields of such intensities should necessarily go beyond a
simple perturbative picture.  The first nonperturbative theory of MPI
was proposed by
\citet{Keldysh64}, \citet{Faisal73} and
\citet{Reiss80}.  Their theory (known as KFR) treated the process
of MPI as a transition of an electron from an initial bound state
into a final state described by the classical Volkov wave
function. The KFR approach provided simple analytical formulas for the
MPI rate which were found in a qualitative agreement with
experiment. Various modifications of the KFR theory were made, in
particular those accounting for the rescattering process
\cite{BLK94,BCL96}.

The KFR theory treated the laser field purely classically. The MPI
problem can also be formulated in an entirely quantum form. The
properties of the scattering matrix in this formulation of the MPI
process have been studied starting from the works of Mower
\cite{Mow66,Mow68}, Gontier et al. \cite{Gont76,Gont81}, \citet{FM81},
\citet{JS82}.  More recently,
\citet{GA88} and Guo, Aberg and Crasemann \cite{GAC89}
put forward an MPI theory (referred hereafter as GAC) which treats
the photoionization process as a QED scattering phenomenon. The emphasis
in this theory was placed on a proper QED description of an electron
interacting with the laser field (the quantum version of the Volkov
states).  Further development of the QED picture of the MPI phenomenon
was made by \citet{GGW00} and \citet{CCL03} who refined the
original GAC theory by including the non-laser modes of the
electromagnetic field.

This fully QED approach, although solving the problem in
principle, was found to be rather difficult to implement in practice
even for simplest atomic targets such as one- or two-electron atoms.
For instance, \citet{CCL03} had to make further drastic
approximations in order to carry out their calculation of MPI on
atomic hydrogen.

One possible solution to this problem is to use some suitable
square-integrable basis to represent the Green functions
occuring in the theory. Such approach, using ideas of the
complex rotation method, has been proposed by \citet{MR83}.
Another solution is to use the field-free atomic
states as building blocks of the theory.  Such a choice is particularly
advantageous when theory is applied to complex atomic systems with
more than one target electron.  Such approach within the context
of the complex rotation method has been used by
\citet{mn1} and \citet{mn2}. Description of
the quantum mechanical evolution of
an atom coupled to a laser field is also possible in the framework
of this method \cite{mn4}.

In the present paper, we outline a quantum formalism for MPI which we
intend to use for practical computations on complex atomic systems.
The formalism is not entirely new and is based on the ideas expressed
earlier in the literature. The main emphasis of this work is on the
pracical implementation of the formlism and turning it into an
efficient computational procedure.  In this development we are
inspired by a series of works by Burke and collaborators who combined
the Floquet description of the laser field with the $R$-matrix
scattering theory. Following the seminal work
\cite{BFJ91}, this approach has been successfully implemented for
calculating the total MPI rate and the level shift in atomic hydrogen
\cite{DTP92}, helium \cite{PDT93}, the negative hydrogen ion
\cite{DPT95} and molecular hydrogen \cite{CGHB01}. Most recently
the $R$-matrix Floquet theory was combined with the basis spline
technique to describe the two-electron MPI from the helium atom in the
ground \cite{FvdH03} and excited states \cite{vdHF01}.  In addition to
the total MPI rate, some differential cross-sections can be also
calculated within the Floquet formalism as was demonstrated for atomic
hydrogen by \citet{PS88} and \citet{Potvliege98} . It is the
most detailed fully differential cross-sections that are of particular
interest to experimentalists and that we intend to evaluate in our
approach.

In the present paper we employ the operator formalism due to
\citet{gw64}. In this formalism, the MPI process is treated as a
decay phenomenon. The partial decay rates and the energy level shifts
are evaluated via the matrix elements of the transition operator which
are found by solving a coupled set of the integral  Lippmann-Schwinger
equations. In this approach the matrix elements of the transition
operator should be taken between the field-free atomic states
accompanied by an integer number of the laser photons.

For one-electron targets, evaluation of the field-free states is
trivial. For two-electron targets, an accurate set of target states,
both discrete and continuous, can be generated by the so-called
convergent close coupling (CCC) method. This method has been
extensively tested for processes with two electrons in the continuum
such as electron scattering on atomic hydrogen \cite{B94} and
low-field double ionization of helium \cite{KB98a,KB98b}. We intend to
use the same set of target states for MPI of He in the
non-perturbative strong-field regime.

In our approach we ignore all the processes of spontaneous emission of
photons, which is justified as we are interested in
processes induced by strong fields. Since the work of \citet{Sh65},
it is known that if one neglects spontaneous processes
in the quantum description of the interaction of laser light and
atom, then the quantum picture and the Floquet
method of solving the time-dependent Schrodinger equation become
equiavalent. Therefore, the present approach and the
ones based on the Floquet anzats (such as \cite{BFJ91}, for
example) are completely equivalent. However, recasting the
MPI description in the form presented below may have some practical
advantages, especially if we have good means of representing
the field-free atomic states for complex atomic systems.
This includes, of course, the states belonging to the continious
spectrum. Such means do exist. As mentioned above, the
CCC method turned out
to be very efficient at representing field-free states
of atomic systems. It was the desire to exploit fully the
possibilities provided by CCC technique that largely motivated us to
give a formulation of the MPI process presented below.

The rest of the paper is organized as follows. In Section~\ref{S2} we
give a formulation of the MPI theory in terms of the field-free atomic
states. In Section~\ref{S4} we consider a model square-well problem,
and in Section \ref{SH} the MPI of hydrogen.  We conclude in
Section~\ref{S5} by outlining a set of problems we intend to consider
in the immediate future.

\section{Nonperturbative framework for the description of the MPI
process.}
\label{S2}

Let us consider a system which consists of a number of photons with a
given frequency $\omega$ and momentum vector $\bm k$ corresponding to
an incident plane-wave, and a target (atom or ion).  We shall describe
the field fully quantum-mechanically and write the Hamiltonian of the
system as
\begin{equation}
\hat H = \hat H_{\rm atom}+\hat H_{\rm field} + \hat H_{\rm int} \ .
\end{equation}
Here $\hat H_{\rm atom}$ and $\hat H_{\rm field}$ have the usual
meaning of the Hamiltonians of the atom and the field:
\begin{eqnarray}
\hat H_{\rm atom}&=&\sum\limits_{i=1}^{N}{{\bm p}_i^2\over 2}
-\sum\limits_{i=1}^{N}{Z\over r_i}
+\sum\limits_{i,j=1,i>j}^{N}{1\over r_{ij}} \nonumber \\
\hat H_{\rm field}&=&\hat N\omega
\label{hams}
\end{eqnarray}
The atomic Hamiltonian is taken in a non-relativistic form. The number
operator $\hat N$ refers to the laser photons only..

The corresponding states of the system consisting of the
non-interacting atom and the field are denoted as $\displaystyle
|\alpha\rangle=|a,n\rangle$, where a set of quantum numbers $a$
defines a state of the atom and $n$ is the number of the laser
photons.  The following notations will be kept throughout the paper:
Greek letters will be used to designate the states of a whole system
``the atom plus external field'', while the Latin letters will be used
for the atomic states. The atomic system of units is in use with
$e=m=\hbar=1$.

The part of the Hamiltonian $ \hat H_{\rm int}$ which describes the
interaction of the atom and the linearly polarized
laser field characterized by
angular frequency $\omega$, wave-vector $\bm k$ and
polarization vector $\bm e$ can be written as
(see e.g. \citet{Sobelman72})
\begin{equation}
\hat H_{\rm int}=-{1\over c}\sum\limits_{i=1}^{N}
\left(\hat {\bm A} \cdot \hat {\bm p}_i-
{\hat{\bm A^2}\over 2 c^2}\right),
\label{hint}
\end{equation}
where $\hat {\bm p}$ is the momentum operator and $\hat {\bm A}$ is a
quantized vector potential normalized to the unit volume:
\begin{equation}
\hat {\bm A}=
\sqrt{2\pi c^2\over \omega}{\bm e}
(\hat a^+_{{\bm k}}e^{i{\bm k}{\bm r}}+
\hat a_{{\bm k}}e^{-i{\bm k}{\bm r}}),
\label{as}
\end{equation}
Here $\hat a^+$ and $\hat a$ have the usual meaning of the operators
creating and annihilating a photon. As mentioned above, we are
interested in sufficiently strong field intencities, when all
processes of spontaneous emission of photons can be neglected.  For
this reason, only one mode corresponding to the wavevector of the
incident laser filed is kept in \Eref{as}.  Therefore, here and
below, $\hat {\bm A}$ describes the laser photons only.  We also
restrict ourselves with the dipole approximation in which the operator
$\hat {\bm A}$ does not act on the atomic coordinates.

The matrix elements of the vector potential operator taken between
the noninteracting states of the system ``atom plus laser photons''
are given by the well-known formulas (see
e.g. \citet{Sobelman72}):
\begin{eqnarray}
\langle a,n|\hat {\bm A} \cdot \hat {\bm a}|b,n-1\rangle &=&
\sqrt{{2\pi c^2 n \over \omega}}\langle a|{\bm e \cdot \hat {\bm a}}|b
\rangle
\nonumber \\
\langle a,n|\hat {\bm A} \cdot \hat {\bm a}|b,n+1\rangle &=&
\sqrt{{2\pi (n+1) c^2\over \omega}}\langle a|{\bm e} \cdot \hat
{\bm a}|b\rangle \ ,
\label{mea}
\end{eqnarray}
where $\hat {\bm a}$ is an arbirtary vector operator acting only on the
coordinates of the atomic subsystem.

In strong fields $n \simeq n-1 \gg 1$ and the coefficients
in \eref{mea}  can be simplified to
\begin{equation}
\sqrt{{2\pi n c^2\over \omega}}\approx
\sqrt{{2\pi (n+1)c^2 \over \omega}}\approx
{Fc\over 2 \omega},
\label{me1}
\end{equation}
where $F$ is the electric field strength related to the energy density
as $F^2=8\pi n \omega$. This leads to the following formulas for the
matrix elements of the operator $\hat H_{\rm int}$:
\begin{eqnarray}
\label{me}
\langle a,n|\hat H_{\rm int}|b,n\pm 1\rangle &=-&
{F \over 2\omega}\langle a|{\bm e}\cdot \hat {\bm p}|b\rangle,
\\
\langle a,n|\hat H_{\rm int}|b,n\pm 2\rangle &=&
{F^2 \over 8\omega^2}\langle a|b\rangle
\nonumber\\
\langle a,n|\hat H_{\rm int}|b,n\rangle &=&
{F^2 \over 4 \omega^2}\langle a|b\rangle
\nonumber
\end{eqnarray}
We shall treat the MPI process as a decay phenomenon within the
framework of the quantum decay theory as described by
\citet{gw64}. We shall be interested in the following process. At
the moment $t=0$ the system ``atom plus external field'' is prepared
in the eigenstate $|\alpha\rangle$ of the Hamiltonian $\hat H_0=\hat
H_{\rm atom}+\hat H_{\rm field}$.  Then interaction $\hat H_{\rm int}$
between atomic and photon subsystems is switched on. Our aim is to
describe possible outcomes of this event.  The partial rates of the
decay of the initial states $|\alpha\rangle$ into various open
channels $|\beta\rangle$ are given by the expressions
\begin{equation}
\Gamma_{\beta}=2\pi |T^{\beta\alpha}(E)|^2
d\rho(E),
\label{tm1}
\end{equation}
where $\rho(E) $ denotes the density of states in the
final state, and the transition operator $T$ satisfies the operator
equation \cite{gw64}
\begin{equation}
\hat T(E)=
\hat H_{\rm int}+\hat H_{\rm int}(1-\hat P_{\alpha})
{1\over E-\hat H_0} \hat T(E),
\label{tm2}
\end{equation}
Here $P_{\alpha}$ is a projection operator on the initial state
$|\alpha\rangle$. In \Eref{tm1} both the matrix element of the
transition operator and the density of states are to be computed at
the energy, corresponding to the
shifted energy of the initial state $|\alpha\rangle$:
$\displaystyle
E=E_{\alpha}+\Delta E_{\alpha}$
Finally, the energy shift $\Delta E_{\alpha}$ of  the
final state is related to the diagonal matrix element of the
transition operator via implicit equation \cite{gw64}
\begin{equation}
\Delta E_{\alpha}={\rm Re} T^{\alpha\alpha}(E)
\label{tm3}
\end{equation}
The form of the latter equation suggests a possiblity of
an iterative scheme for the determination of the level shift.
Equations \eref{tm1}-\eref{tm3} provide the basis of our calculation
scheme for evaluating the rates of various multiphoton processes.

In the alternative formalism \cite{PS88}, when the MPI process is
treated as a scattering phenomenon, the expression for the
$T$-operator must be corrected for the virtual transitions between
various photodetachment channels. This correction comes from a careful
examination of the boundary conditions which must be imposed on a
scattering wave function in the Floquet representation. In its
simplest form, this can be achieved by dividing the conventional
expression for the matrix element of the $T$-operator by a Bessel
function $J_0$
\cite{PS88}.  The problem outlined  by Eqs.~(\ref{tm1}) -(\ref{tm3})
is a decay, or an initial state, problem.  As such, it requires an
accurate evaluation of the initial state which subsequently decays
into various open channels. Indeed, \Eref{tm1} is a result of
evaluation of the quantum mechanical amplitude $\displaystyle
\langle \beta|\Psi(t)\rangle$  where $\Psi(t)$ evolves from the
state $|\alpha\rangle$ at $t=0$ \cite{gw64}. Therefore no additional
correction to the transition amplitudes \eref{tm2} is needed.

By introducing in Eq.~(\ref{tm2}) a complete set of states of $\hat
H_0$, this equation can be rewritten in the form of a spectral
representation:
\begin{equation}
T^{\beta \alpha}=\hat H_{\rm int}^{\beta \alpha}+
\isum\limits_{ \gamma\neq\alpha }
{\hat H_{\rm
int}^{\beta \gamma} \  T^{\gamma \alpha}\over
E_{\alpha}+\Delta E_{\alpha}-E_{\gamma}+i\epsilon}
\label{ls4}
\end{equation}
Here the sign of $i\epsilon$ gives the rule of bypassing the pole when
performing the integration over the continuum spectrum.

The shift of the energy of the initial state is explicitely
included into the set of Eqs.(\ref{ls4}). This circumstance
may be especially important for the atomic systems with
more that one electron. When the escaping electron leaves
the atomic system, the remaining ion still interacts with the
electromagnetic field, the energy shift in the
Eqs.(\ref{ls4}) takes into account effects of this interaction.

\begin{figure}[h]
\epsfxsize = 16cm
\epsffile{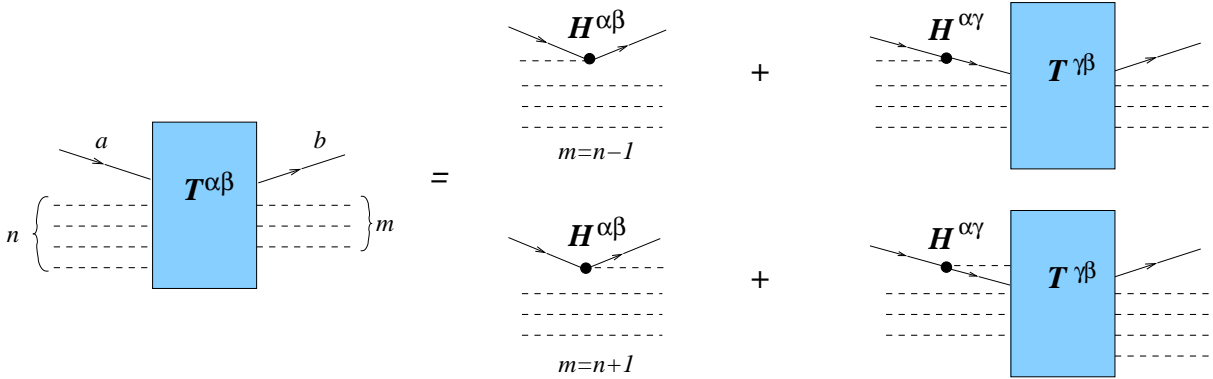}
\caption{\label{fig1}
Diagram representation of the Lippmann-Schwinger equation for
multiphoton ionization.  The graphical symbols are described in the
text.}
\end{figure}

In Fig. \ref{fig1} we give a graphical representation of \Eref{ls4}. Here a
straight line with an arrow to the right represents a target atom and
the dashed lines are used to depict the photons. A vertex with two
electron lines and one photon line represents a dipole matrix element
\eref{me} which incorporates many-electron correlation in the
target. For simplicity we do not show the quadrupole and monopole
matrix elements \eref{ls4} which are parametrically small for not very
strong fields $F<1$. A rectangular block stands for the $T$-matrix
\eref{ls4}. In the low-field regime the integral term in the
right-hand side of \eref{ls4} can be ignored and the atomic ionization
is described by the bare matrix element $\langle a,n|\hat H_{\rm
int}|b,n\pm 1\rangle $. The strong field effects are incorporated in
the integral term and include multiple absorption and emission of the
laser photons.

The sum over the spectrum of $\hat H_0$ in \eref{ls4} includes
summation over various number of photons as well as summation over
bound atomic states and integration over the continuous spectrum of
the atom. Computation of the integral in \eref{ls4} can be greatly
simplified by introducing a discrete set of target pseudostates which
provides an adequate quadrature rule.  As the result of such a
discretization the set of equations \eref{ls4} becomes a linear system
on the unknown elements of the $T$-matrix. Once this linear system is
solved, all information about the integral and differential features
of the MPI process can be obtained from the matrix elements of the
$T$-operator taken at the energy corresponding to the shifted energy
of the final state as in Eq.~(\ref{tm1}) (the so-called on-shell
matrix elements). We note that in the weak field limit summation in
\eref{ls4} can be restricted to the atomic variables and we arrive to
the expression for the dipole matrix element given by Eq.~(12) of
\citet{KB98a}.

\subsection{Square well model}
\label{S4}

To illustrate feasibility of our approach we first
apply it to a model problem of an MPI process
from a one-dimensional square well. It is this model system that was
considered by \citet{BFJ91} in their seminal paper which gave
rise to the spectacular success of the $R$-matrix Floquet theory.  We
consider here an electron bound initially in a square well potential
$V=-2.5$ a.u. for $0<x<1$ and $V=0$ for $x > 1 $, with the boundary
condition $R(0)=0$ imposed on the wave functions.  This potential
supports only one bound state $a$ with an energy $E=-0.4657$ a.u. We
consider an MPI process when the electric field with the frequency
$\omega=0.2$ is applied such that at least three photons are needed to
ionize the system. To solve this problem we follow the steps outlined
in Sec.~\ref{S2}. The set of equations \eref{ls4} is converted into a
linear system on the $T$-matrix elements by choosing a suitable
discretization procedure for the continuous spectrum integration.  The
summation over a set of intermediate states $\gamma$ in the
Eq.~(\ref{ls4}) is a sum over various numbers of photons $n_{\gamma}$
and summation and integration over the set of variables specifying the
field-free electron states. For the square well model the latter are
determined by a single quantum number (momentum of the states
belonging to the continuous spectrum or energy of the discrete
spectrum). For a given number of photons in the set of intermediate
states $\gamma$ we first determine if there is a pole in the integral
over the continuous spectrum. Similarly to electron scattering
calculations of \citet{B94}, we divide the momentum integration
interval $(0,\infty)$ into five subintervals.

If the pole is present in the integral, the first two subintervals are
chosen to be $(0,k_{\rm pole})$ and $(k_{\rm pole},2k_{\rm pole})$
with a typical number of 60 momentum points in each interval.  Then
the delta-function singularity is isolated and the remaining principle
value integral is evaluated by a modified Simpson's rule
\cite{B94}. The remaining part of the momentum integral is divided as
follows: $(2k_{\rm pole },4)$ (50 integration points), $(4,10)$ (40
points) and $(10,350)$ (40 points).  These intervals are pole-free and
the integration is performed by using a conventional Simpson's
rule. If there is no pole in a given channel the integration procedure
remains essentially the same except for the boundary of the first and
second intervals which is chosen at one atomic unit of momentum.  In
this case the conventional Simpson's rule is used throughout.

We retained various numbers of photons in the intermediate state of
\Eref{ls4}. For simplicity, we count this number from the baseline of
three photons in the initial state which we denote as
$\alpha=|a,3\rangle$. In the intermediate and final states we can then
have negative numbers of photons. For example, $n=-1$ in the final
state $|\gamma\rangle$ would mean that four photons have been
absorbed.  Using this convention, the calculations we performed can be
denoted as ($n^{\rm min},n^{\rm max}$) meaning that the number of
photons in the intermediate and final states ranges from $n^{\rm min}$
to $n^{\rm max}$.

The problem which arises
immediately when one tries to apply Eqs.(\ref{ls4}) for the
square-well model is the singularity of the matrix elements $\hat
H^{\beta\alpha}_{\rm int}$ when both states $\alpha, \beta$ lie in the
continuum. This is, of course, a consequence of
the choice of the gauge of the vector potential in
formulas \eref{hint}. Fortunately, as we shall show in
the section, this problem can be avoided for real atomic
systems by choosing the interaction Hamiltonian
in the Kramers-Hennerberger (acceleration)
form. For the square-well potential such
choice of the interaction Hamiltonian is not possible
due to the singular character of the potential.
We shall have, therefore, to devise a suitable
procedure of the regularization of the matrix elements
of operator \eref{hint} when both states belong to
continious spectrum. There is vast amount of literature
devoted to the study of such matrix elements for
various potentials (e.g.\cite{VP90,Tr89,Kor97}.
We dealt with this problem along the lines
described in the paper by \citet{mn4}.
One can show that for any potential for which the
asymptotic behavior of the continuous wave function is given by
$\displaystyle R_k\propto
\sin{(kr+\delta)}$, the following result holds:
\begin{equation}
\int\limits_0^\infty R_{k_i}(r){d\over dr}R_{k_f}(r)\ dr=
V(k_i,k_f)+{\cos{(\delta_i-\delta_f)}\over 2}  P{1\over k_i-k_f},
\label{pv}
\end{equation}

Here $V(k_i,k_f)$ is a regular function and symbol $P$ has a usual
meaning of the principal value integral.  Note that this divergence is
present only when the velocity form (as in the present paper) or the
length form are used for the electromagnetic interaction operator, it
is absent if the acceleration form is used. Presence of the
singularity in the matrix elements in the velocity and length gauges
is a consequence of the use of the spectral representation for the
Green function in \eref{ls4} and should disappear from the final
result after the summation over the whole spectrum is carried out. A
convenient way to deal with such integrals is to introduce a suitable
regularization procedure.  Different procedures of this sort can be
devised. We shall illustrate the use of the two simplest methods.

The first regularization scheme consists in computing all the
divergent integrals over some finite interval $(0,D)$, hoping that
final results depend only very weakly upon $D$ if its value is
suitably chosen. That this is indeed so can be seen from the results
for the level shift presented in \eref{tab0}. According to
Eq.~(\ref{tm3}), this quantity is related to
the real part of the diagonal matrix
element of the $T$-operator.

To determine the shift we devised
a simple iterative procedure, based on the Eq.~(\ref{tm3}).
The linear
system \eref{ls4} was solved with a trial value
$\Delta E_{\alpha}=0$, than the value given by the r.h.s of
the Eq.~(\ref{tm3}) was adopted as a new trial value
for $\Delta E_{\alpha}$, the
procedure was repeated untill convergence was achieved. Typically,
three iterations were sufficient to achieve
convergence on the level of a fraction of a percent.

The results of a (-4,4) calculation using this
strategy are shown in the Table \ref{tab0}.
The calculation was performed for different values
of the cutoff parameter $D$. As one can see, the
values for the level shift are fairly stable with respect to the
variations of the regularization parameter $D$. The stability of the
results with $D$ becomes somewhat worse for very strong fields
($F=0.2$~a.u. in the Table \ref{tab0}). This, however, is to be expected as
for such strong fields the (-4,4) calculation is probably too small
and a larger number of photons should be included to obtain more
accurate results.  Results in the Table
\ref{tab0} should be compared with the
plot illustrating the level shift as a function of the electric field
strength from the time-dependent $R$-matrix calculation of
\citet{BB97}. Although the numerical values are not reported in
this paper, a visual comparison of our data with the the plot quite
satisfactory.

\begin{table}[h]
\caption{\label{tab0}
Dependence of the level shift from the (-4,4)
calculation upon the regularization parameter $D$. }
\begin{ruledtabular}
\begin{tabular}{lllll}
$D$  & $F=0.06$~a.u. & $F=0.1$~a.u. & $F=0.14$~a.u. & $F=0.2$~a.u.\\
\hline
\noalign{\smallskip}
  30 & -0.023307 & -0.065753  & -0.13132 & -0.28124  \\
  40 & -0.023305 & -0.065704  & -0.13045 & -0.27645  \\
  50 & -0.023311 & -0.065794  & -0.13245 & -0.29102  \\
  60 & -0.023316 & -0.065743  & -0.13386 & -0.29239  \\
  70 & -0.023313 & -0.065749  & -0.13193 & -0.29839  \\
\noalign{\smallskip}
\end{tabular}
\end{ruledtabular}
\end{table}

Another regularization scheme, which we found to give somewhat more
stable results for the ionization rates is based on the regularization
formula for a principal value integral:
\begin{equation}
P{1\over k_i-k_f} =\lim_{\epsilon\to 0}
{k_i-k_f\over \epsilon^2+(k_i-k_f)^2}
\label{pv1}
\end{equation}
Again, one can hope that if the regularization is properly
implemented, the final results do not depend upon the regularization
parameter $\epsilon$.  We analyze these results in the form of the
partial ionization rates computed according to Eq.~(\ref{tm1}). The
momentum $k_f$ of the final state is determined via the energy of the
initial state $\alpha$, number of photons in the final state $n_f$,
and the energy shift $\Delta E$, given by Eq.~(\ref{tm3}):
\begin{equation}
{k_f^2\over 2}=E_{\alpha}+\Delta E-n_f\omega
\label{kf}
\end{equation}
The total ionization rate is the sum of the partial rates over all the
open channels.

\begin{table}[h]
\caption{\label{tab1}
Convergence of the partial ionization rates $\Gamma_3$
$\Gamma_4$, $\Gamma_5$ from the (-3,3)
calculation with respect to the regularization parameter $\epsilon$ in
\ref{pv1}. The field strength is $F=0.1$ a.u. }
\begin{ruledtabular}
\begin{tabular}{llll}
$\epsilon \times 10^2$ & $\Gamma_3$ ($10^{-4}$~a.u.) &
$\Gamma_4$ ($10^{-6}$~a.u.) & $\Gamma_5$ ($10^{-6}$~a.u.) \\
\hline
\noalign{\smallskip}
  100    &1.582 & 1.7958 & 5.584    \\
  50     &1.829 & 0.9439 & 5.866    \\
  25     &2.045 & 0.2198 & 6.165    \\
  12.5   &2.216 & 0.1252 & 6.518    \\
  6.25   &2.334 & 0.4667 & 6.552    \\
  3.125  &2.404 & 0.7188 & 6.529    \\
  1.563  &2.441 & 0.8496 & 6.547    \\
  0.781  &2.454 & 0.9187 & 6.578    \\
  0.390  &2.456 & 0.9537 & 6.610    \\
  0.195  &2.460 & 0.9682 & 6.622    \\
\noalign{\smallskip}
\end{tabular}
\end{ruledtabular}
\end{table}

In the Table \ref{tab1} we show the partial ionization rates
$\Gamma_3$,$\Gamma_4$,$\Gamma_5$ from our (-3,3) calculation. The data
presented in the Table clearly indicate convergence as $\epsilon\to
0$. In the Table \ref{tab2} we present the partial ionization rates $\Gamma_i$
where $i$ has a meaning of the number of absorbed photons for the
field strength $F=0.1$~a.u.  We used different number of laser photons
($n^{\rm min},n^{\rm max}$), with $n^{\rm max}=3,4$ and $n^{\rm
min}=-4,-3,-2,-1,0$.  In these calculations the number of open
channels may differ. For example, in the (-4,3) calculation one may
have the final states with three, four, five, six or seven photons
absorbed. We remind the reader that we count the number of photons
from the baseline of three photons in the initial state.

\begin{table}[h]
\caption{\label{tab2}
Ionization rates for $F=0.1$~a.u.}
\begin{ruledtabular}
\begin{tabular}{cccc}
$n^{\rm min},n^{\rm max}$
& $\Gamma_3$ ($10^{-4}$~a.u.)
& $\Gamma_4$ ($10^{-6}$~a.u.)
& $\Gamma_5$ ($10^{-6}$~a.u.) \\
\hline
\noalign{\smallskip}
-2,3 & 2.4586 & 1.0628 & 6.7172 \\
-3,3 & 2.4562 & 0.9537 & 6.6101 \\
-4,3 & 2.4569 & 0.9644 & 6.5025 \\
-2,4 & 2.3228 & 0.8676 & 6.6303 \\
-3,4 & 2.3208 & 0.7817 & 6.5132 \\
\noalign{\smallskip}
\end{tabular}
\end{ruledtabular}
\end{table}

As one can see, the partial rates corresponding to processes with
large number of photons absorbed (four and five) show certain
stability with respect to the number of photons included in the
calculation, which gives us confidence in the results obtained for
these partial rates. One may note, in particular, that for
$F=0.1$~a.u. the partial rate $\Gamma_5$ becomes larger than the rate
$\Gamma_4$.

\begin{figure}[h]
\includegraphics[height=7cm]{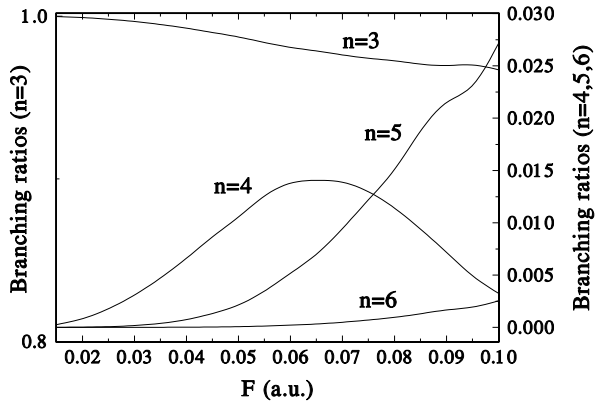}
\caption{\label{fig2}
Branching ratios of the partial ionization rate to the total rate for
various open channels in the square well model.}
\end{figure}

The partial contribution of various open channels to the total
ionization rate can be judged from the data presented in the Fig.
\ref{fig2} where we plot the branching ratios for the channels with
various numbers of photons in the final state. Again, one can see
that for not very large field strengths ( $F<0.05$~a.u.)  the
contributions of the channels with more than three photons absorbed
(labeled $n=4$ and $n=5$ in the Figure) can be neglected, though
for larger fields their effect is becoming increasingly important.

Incidentally, one may note, that even (0,3)
calculation would not correspond exactly to the third order
perturbation theory. Indeed, solution of the coupled set of equations
\eref{ls4} amounts effectively to summation of an infinite subset of
the perturbation theory terms as is illustrated in
the Fig. \ref{fig1}.

The situation becomes somewhat more complicated for stronger fields
($F>0.1$~a.u.) where the level shift is so large that the channel
closing may occur. A look at the data for the shift presented in the
Table \ref{tab0} shows that when $F\approx 0.14$~a.u. the energy of the
initial bound state $\alpha$ is: $E\approx -0.6$~a.u. and absorption
of the three photons is no longer sufficient to ionize the system. A
more precise value of the field for which the channel corresponding to
absorption of the three photons closes is $F\approx 0.1435$~a.u.  as
shown by \citet{BB97}. Rich variety of phenomena may occur when
field approaches this critical value (and larger ones, corresponding
to closing of other channels), in particular ionization rates as
functions of electric field intensity may have discontinuities
\cite{BB97}.

\begin{figure}[h]
\includegraphics[height=7cm]{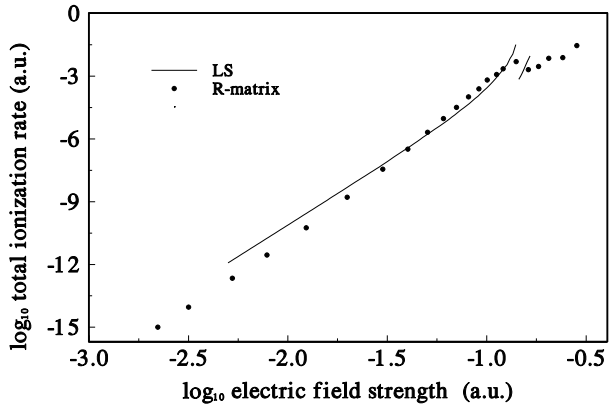}
\caption{\label{fig3}
Total ionization rate as a function of the field strength. The solid
line shows the present (-4,3) calculation. Results of the
time-dependent $R$-matrix calculation of \citet{BB97} are shown
as points.  }
\end{figure}

In the Fig. \ref{fig3}, we plot the total ionization rate as a function of the
field strength from the (-4,3) calculation.  The same data for the
total ionization rate are presented in the Table \ref{tab3}.  In the figure
we make a comparison with the time-dependent R-matrix calculation by
\cite{BB97} which was found to be almost identical to the earlier
Floquet R-matrix results by \citet{BFJ91}.


Our data show certain trend to deviate from those obtained in
\citet{BB97} and \citet{BFJ91} for smaller field
intensities, around $F=0.02$ a.u. This deviation could probably be
attributed to difficulties of numerical character which are due to the
necessity to manipulate the divergent integrals. These are  most
difficult to cope with for smaller fields when rates are small. This
difficulties may easily be avoided for 'real world' problems. For any
realistic atomic system, writing the electromagnetic interaction
operator in the acceleration form would make all the integrals
well-defined and convergent.

\begin{table}[h]
\caption{\label{tab3}
Total ionization rate from (-4,3) calculation.}
\bigskip
\begin{ruledtabular}
\begin{tabular}{ll ll}
$F$(a.u.)& $\Gamma$(a.u.) &
$F$(a.u.)& $\Gamma$(a.u.) \\
\hline
\noalign{\smallskip}
0.025 & 0.192$\times 10^{-7}$ & 0.14  & 0.237$\times 10^{-1}$ \\
0.05  & 0.135$\times 10^{-5}$ & 0.15  & 0.119$\times 10^{-2}$ \\
0.075 & 0.291$\times 10^{-4}$ & 0.16  & 0.358$\times 10^{-2}$ \\
0.1   & 0.254$\times 10^{-4}$ & 0.175 & 0.356$\times 10^{-1}$ \\
0.125 & 0.279$\times 10^{-2}$ \\
\noalign{\smallskip}
\end{tabular}
\end{ruledtabular}
\end{table}

\subsection{Hydrogen atom}
\label{SH}

We turn now to a more realistic system -- the  hydrogen atom in
the presence of a monochromatic linearly-polarized laser field.
As we mentioned above, this system is in some sense
simpler than the model square-well problem, as we
can use the Kramers-Hennebeger form of the interaction
Hamiltonian \cite{kh1,kh2}, which,
as we shall see, allows to avoid all the problems of
divergence of the matrix elements.

If electromagnetic field is treated classically,
the Kramers-Henneberger Hamiltonian can be obtained from the
minimal-coupling Hamiltonian $\displaystyle \hat H_{\rm min}$ given
by Eqs.~(\ref{hams}-\ref{hint}) by
means of a canonical transformation
$\displaystyle \hat H_{\rm KH}=
e^{i\hat T}\hat H_{\rm min}e^{-i\hat T}$
generated by the
operator \cite{kh2}:
\begin{equation}
\hat T=-{1\over c}\int\limits_0^t {\bm A}(\tau){\bm p}\ d\tau+
       {1\over 2c^2}\int\limits_0^t {\bm A}^2(\tau)\ d\tau \ ,
\label{khf}
\end{equation}
where ${\bm A}$ is the vector potential and ${\bm p}$ is the momentum
operator.

In the present approach, when electromagnetic field
is described quantum-mechanically, we need a quantum
analog of the transformation (\ref{khf}). Such transformation is
known as the Pauli-Fierz canonical transformation \cite{PF}.
For the case of a sufficiently intense linearly polarized monochromatic
light, when all the processes of
spontaneous radiation are neglected
and the quantized vector potential is given by the Eq.(\ref{as}),
with only the laser mode photons preserved in the expansion,
this transformation assumes the form \cite{lac}:

\begin{eqnarray}
{\bm r}&=&{\bm r}'+ {{\bm a}\over \omega} \hat Q \nonumber \\
{\bm p}&=&{\bm p}' \nonumber \\
\hat Q&=&\hat Q' \nonumber \\
\hat P&=&\hat P'-{{\bm a}{\bm p}'\over \omega},
\label{can}
\end{eqnarray}

where $\displaystyle {\bm a}=2{\bm e}\sqrt{\pi\over\omega}$,
${\bm e}$ is polarization vector, operators $\hat Q'$ and $\hat P'$
are expressed in terms of creation and annihilation operators
in Eq.(\ref{as})) via:

\begin{eqnarray}
\hat a_{\bm k}&=&{1\over \sqrt{2}}\left(\hat P+i\hat Q\right)\nonumber \\
\hat a^+_{\bm k}&=&{1\over \sqrt{2}}\left(\hat P-i\hat Q\right)
\label{qp}
\end{eqnarray}

In Eqs.(\ref{can}),(\ref{qp}) primed quantities refer to
original (laboratory) frame, and non-primed to the
Kramers-Henneberger frame.

Under the transformation (\ref{can}) the minimal-coupling
Hamiltonian (\ref{hams}) becomes:
\begin{equation}
\hat H_{\rm KH}={{\bm p}^2\over 2}-{1\over r}+\hat H_{\rm field}+
\hat H_{\rm intKH} \ ,
\label{hams1}
\end{equation}
where $\hat H_{\rm field}$ has the same meaning as in
Eq.~(\ref{hams}) and the
interaction Hamiltonian has the form:
\begin{equation}
\hat H_{\rm intKH}=
-{1\over |\bm r+\hat{\bm \alpha}|}+{1\over r} \ ,
\label{hint1}
\end{equation}
where the operator $ \hat{\bm \alpha}=\hat{\bm F}/ \omega^2$ is
related to the operator of the electric field intensity $\hat{\bm F}$.

Certain amount of care should be taken when
physically relevant information is to be extracted
from the transformed Hamiltonian $\hat H_{\rm KH}$,
in particular, when our aim is to study time-evolution
of a quantum system prepared at the moment $t=0$ in
a given eigenstate of the field-free Hamiltonian.
The Kramers-Henneberger transformation consists,
basically, in transforming the Schrodinger equation
into a non-inertial frame moving with electron oscillating in the
field (in the quantum version of this transformation this
can be seen from the first of Eqs.(\ref{can}), the quantity on
the r.h.s of this equation being essentially an electric field
operator).
Both the initial and final state vectors must then be transformed
accordingly \cite{VK,RB}. This circumstance may introduce additional
complexity in the problem. 
Below, we report results on the total 
ionization rate for multiphoton ionization of hydrogen atom. 
One can present the following argument showing that if we are 
interested in integral characteristics, such as total ionization 
rate, one need not transform initial and final states. Indeed, 
solution of the set of equations (\ref{ls4}) with the 
Kramers-Henneberger Hamiltonian (\ref{hams1}),(\ref{hint1}) 
is essentially a study of the singular points of the resolvent 
operator $\displaystyle (E-\hat H_{\rm {KH}})^{-1}$ (we will be 
looking, of course, for the points lying on the unphysical sheets 
of the Riehmann surface). 
Total ionization rate from the ground 
state of hydrogen obtained below from such a study of the Hamiltonian 
$\hat H_{\rm KH}$ is in fact related to an 
imaginary part of one such singular point. On the other hand, 
total ionization rate obtained using the minimal-coupling 
Hamiltonian is a singular point of the resolvent operator
$\displaystyle (E-\hat H_{\rm {min}})^{-1}$ which is
connected to the resolvent operator 
$\displaystyle (E-\hat H_{\rm {KH}})^{-1}$ by means of the 
canonical transformation (\ref{can}). The latter transformation
does not change position of singular points, hence, study
of singular points $H_{\rm KH}$ alone (by means of solving
set of Eqs.(\ref{ls4})) should give us correct value for
the total ionization rate.

To apply the strategy based on Eqs.(\ref{ls4}) to the
system described by the Kramers-Hennebeger Hamiltonian, one should be able
to compute the matrix elements of the operator
\eref{hint1}. The procedure which we devised for this purpose
is described below.

\subsubsection{Calculation of matrix elements of
$\hat H_{\rm int}$.}

We shall need the following formula for the matrix elements
of the operator of the electromagnetic field, which can be obtained
analogously to the Eqs.(\ref{mea}):
\begin{eqnarray}
\label{mee}
\langle a,n|\bm a \hat{\bm F}|b,m\rangle &\approx&
{1\over 2}
\langle a|\bm a \bm F|b \rangle \qquad n\to\infty, |m-n|=1 \nonumber \\
\langle a,n|\bm a \hat{\bm F}|b,m\rangle &=& 0 \qquad |m-n|\neq 1
\end{eqnarray}
In Eq.~(\ref{mee}) ${\bm F}$ is the classical vector of electric field
strength, $\bm a$ is an arbitrary vector operator not acting on the
photon variables.
The correction term to the non-zero matrix element
in Eq.~(\ref{mee}) is of the order of $n^{-{1\over 2}}$ and is negligible in
the limit of large intensities which is of interest to us in the
present paper.

Consider first the matrix element
\begin{equation}
M_{mp}=\langle n+p|(\bm a \hat{\bm F})^m|n\rangle
\label{ma1}
\end{equation}

We are interested now only in photonic variables, therefore we omitted
in \eref{ma1} any reference to atomic variables.  It is easy to see
that $M_{mp}\neq 0$ only if $m=p+2k$ with an integer nonnegative
$k$. If this condition is satisfied, it is easy to show using
Eqs.(\ref{me}) and simple combinatorial analysis that
\begin{equation}
M_{mp}=\left({\bm a \bm F\over 2}\right)^m
{m!\over
\left({m+p\over 2}\right)!
\left({m-p\over 2}\right)!}
\label{ma2}
\end{equation}
We can now compute a more complicated matrix element
\begin{equation}
\langle n+p|\exp{\{\bm a \hat{\bm F}\}}|n\rangle
\label{ma3}
\end{equation}
By expanding the exponential function in the above expression, with
the use of Eq.~(\ref{ma1}) and known series expansion for the
Bessel function $I_p(x)$ \cite{abr}

\begin{equation}
I_p(x)=\left({x\over 2}\right)^p \sum\limits_{k=0}^\infty
{\left({x\over 2}\right)^{2k}\over k!\Gamma(k+p+1)}
\label{bes}
\end{equation}

we can obtain the following expression:
\begin{equation}
\langle n+p|\exp{\{\bm a \hat{\bm F}\}}|n\rangle=
I_p(\bm a\bm F),
\label{ma4}
\end{equation}
where $I_p(x)$ is a modified Bessel function of order $p$ \cite{abr}.

We may now turn to the computation of the matrix elements of the
operator \eref{hint}.  Representing \eref{hint} as a Fourier
transform:
\begin{equation}
{1\over |\bm r_i+\hat{\bm \alpha}|}=
{1\over 2\pi^2}\int {e^{i\bm q(\bm r+\hat{\bm \alpha})}\over q^2}\ d\bm q,
\label{ma5}
\end{equation}
using expression \eref{ma4} and known integral
representation for the modified Bessel function $I_p(x)$ \cite{abr}:

\begin{equation}
I_p(x)=
{1\over \pi}\int\limits_0^{\pi}
e^{x\cos{\theta}}\cos{p\theta}\ d\theta
\label{bes1}
\end{equation}
one can obtain the following formula:
\begin{equation}
\left\langle n+p\left|
{1\over r_i}-
{1\over |\bm r_i+\hat{\bm F}/ \omega^2|}
\right|n\right\rangle=
{1\over \pi}\int\limits_0^{\pi}
\cos{p\theta}
\left({1\over r_i}-{1\over |\bm r_i+
{\bm F}\cos{\theta}/ \omega^2|}\right)
\ d\theta
\label{ma6}
\end{equation}
To proceed further, we use a spherical harmonics expansion
\cite{Varshalovich}:
\begin{equation}
{1\over \Big|\bm r_i+\bm F\cos\theta/ \omega^2\Big|}=
\sum\limits_{k=0} \sqrt{4\pi\over 2k+1}{r_<^k\over r_>^{k+1}}
[-{\rm sign}(\cos{\theta})]^k \ Y_{k0}({\bm r}),
\label{ma7}
\end{equation}
where $r_<$ ($r_>$) is the smaller (greater) of $r_i$ and $F\cos\theta/
\omega^2$.  Here the field ${\bm F}$ is directed along the $z$-axis.
\Eref{ma7} allows separation of the  radial and angular
variables.
Angular parts are evaluated analytically using integrals of
products of several spherical functions \cite{Varshalovich}:

\begin{eqnarray}
\int Y_{l_1m_1}({\bm \Omega})
Y_{l_2m_2}({\bm \Omega})Y_{l_3m_3}({\bm \Omega})\ d{\bm \Omega}
=\nonumber \\
\sqrt{
(2l_1+1)(2l_2+1)(2l_3+1)\over 4\pi}
\left(\matrix{l_1 & l_2 & l_3 \cr 0 & 0 & 0}\right)
\left(\matrix{l_1 & l_2 & l_3 \cr m_1 & m_2 & m_3}\right)
\label{var}
\end{eqnarray}

Radial integrals are computed numerically.

\subsubsection{Numerical Results}
\label{SH3}

We first illustrate our procedure of the calculation of the matrix
elements of the interaction Hamiltonian. With the help of the formulas
\eref{ma6} and\eref{ma7} we calculated the integrals $\displaystyle
\langle a,n+p|\hat H_{\rm int}| b,n\rangle$.  In the case of hydrogen
atom the atomic states $|a\rangle$ are characterized by energies and
angular momenta. We computed integrals for sufficiently dense grid of
energies. As for the range of momenta and a number of photons retained
in the calculation, their upper limits depend on the value of the
electric field strength. For moderately high field strengths
($F\approx 0.1$~a.u.)  it was sufficient to compute integrals with
$l<l_{\rm max}=3$ and $p<p_{\rm max}=7$. The integrals were computed
and stored on a disk.

As one can see from the Eq.~(\ref{ma6}),
in the weak field limit $p=1$ and the
interaction Hamiltonian in the Kramers-Hennerberger form reduces to
the operator $\displaystyle {\hat{\bm F}{\bm r}\over r^3\omega^2}$
which is commonly used in the first order calculations performed in
the acceleration gauge. The parameter which measures the departure of
the matrix element of operator \eref{ma6} from the first order result
is the ratio $F/\omega^2$. This departure can be quite significant. To
give an illustration of the relative role of the higher order
corrections in Eq.~(\ref{ma6}) we present in Table \ref{tab1h}
few values of
the matrix elements $\displaystyle \langle 1s,n=1|\hat H_{\rm int}|
2p,n=0\rangle$ and $\displaystyle \langle 1s,n=1|\hat H_{\rm int}|
kp,n=0\rangle$ with momentum $k=1$~a.u. for different frequencies and
electric field strength of $F=0.0534$~a.u. First order matrix elements
are also given in the Table \ref{tab1h}
(marked as PT). As can be seen, for small
frequencies the deviation of the matrix elements from the first order
values can be significant.

\begin{table}[h]
\caption{\label{tab1h}
Matrix elements of the operator \eref{hint}) for F=0.0534 a.u.}
\begin{ruledtabular}
\begin{tabular}{l cc cc}
Angular &
\multicolumn{2}{c}{$\displaystyle \langle 1s,n=1|\hat H_{\rm int}| 2p,n=0\rangle$} &
\multicolumn{2}{c}{$\displaystyle \langle 1s,n=1|\hat H_{\rm int}| kp,n=0\rangle$} \\
frequency & Present & PT & Present & PT\\
\hline
\noalign{\smallskip}
0.65 &  0.03070 &  0.03118 & 0.02126  & 0.02143 \\
\noalign{\smallskip}
\noalign{\smallskip}
0.184 & 0.04678 & 0.08261 & 0.11335 & 0.26796 \\
\noalign{\smallskip}
\end{tabular}
\end{ruledtabular}
\end{table}

To facilitate comparison with the literature, we performed a series of
calculations for the same set of laser field strengths and frequencies
as reported by \citet{DPT95} who used a combination of the
Floquet ansatz and the $R$-matrix methods.  Our results for the level
shift and total ionization rate together with the literature values
are summarized in the Table \ref{tab2h}.

When presenting their data for the level shifts, \citet{DPT95}
who performed their calculation in the length gauge subtracted the
ponderomotive energy $E_P=F^2/4\omega^2$ from the total shift. Since
the level shifts in the acceleration and the length gauge are related
as $\displaystyle \Delta E^A=\Delta E^L- E_P$ \cite{BFJ91} the
two sets of data are directly comparable.

\begin{table}[h]
\caption{\label{tab2h}
Total ionization rates and shifts for atomic hydrogen in a linearly
polarized field of frequency $\omega$ and strength F (a.u.). Numbers
in parenthesis represent data of \citet{DPT95}}
\begin{ruledtabular}
\begin{tabular}{c c ll ll}
Angular  & Field  &
\multicolumn{2}{c}{Total ionization rate} &
\multicolumn{2}{c}{Shift~~~~~} \\
frequency & strength& Present& \citet{DPT95}&Present& \citet{DPT95}\\
\hline
\noalign{\smallskip}
0.65 &  0.0534 & 0.00263  & 0.00256 & 0.000364 & 0.000360\\
\noalign{\smallskip}
     &  0.754  & 0.13 & 0.14 &   0.176 & 0.195 \\
\noalign{\smallskip}
\noalign{\smallskip}
0.184& 0.0169  & $9.2\times 10^{-6}$ & $8.8\times 10^{-6}$& -0.002910  &-0.002543\\
\noalign{\smallskip}
     & 0.0534  & $1.33\times 10^{-3}$ & $1.40\times 10^{-3}$ & -0.0280 & -0.0257 \\
\noalign{\smallskip}
\end{tabular}
\end{ruledtabular}
\end{table}

We consider first the case of the laser field with $\omega=0.65$~a.u.
and $F=0.0534$~a.u. As noted by \citet{DPT95}, this is still a
perturbative regime. In Eqs.(\ref{ls4})
we need not, therefore, retain very
large set of intermediate states $\gamma$ to achieve convergence.

In the sum over the intermediate states we included the hydrogen atom
states with orbital momentum $l=0,1$ and a set of photonic states with
numbers of photons ranging from $n_{\rm min}=-2$ to $n_{\rm max}=3$.
Counting the photons we use here the same convention we used in the
previous section, when dealing with the square well potential model.
We count the total number of photons in the system from the minimal
number of photons needed to ionize the initial state. Thus, for the
angular frequency $\omega=0.65$ and the ground state of the hydrogen
atom, we have one photon in the initial state.  The state with $n=-1$
will then correspond to the state in which two photons have been
absorbed.

As one can see from the Table \ref{tab2h},
we achieve quite a good agreement
with the data of \citet{DPT95}, both for the total rate and level
shift.  For much larger field and the same frequency
$\omega=0.65$~a.u.  we need a larger set of intermediate states to be
included Eq.~(\ref{ls4}). Convergence was achieved when we included
the hydrogen atom states with $l<l_{\rm max}=4$ and photonic states
with numbers of photons ranging from $n_{\rm min}=-2$ to $n_{\rm
max}=3$.

For the process of genuine multiphoton ionization ($\omega=0.184$) we
also have reasonably good agreement with the results of the Floquet
R-matrix calculation \cite{DPT95}.  For the field strength $F=0.0169$
we used atomic states with angular momenta $l<4$ and photonic states
with $n_{\rm min}=-1$ and $n_{\rm max}=5$. This field value is still
in the perturbative regime.

Truly nonperturbative effects appear for larger field strengths. To
obtain the values of the shift and total rate for $F=0.0534$ we used
atomic states with $l<4$ and photonic states with $n_{\rm min}=-2$ and
$n_{\rm max}=5$.

\section{Conclusion}
\label{S5}

We developed a non-perturbative formalism which describes the MPI
process by a set of coupled integral equations of the
Lippmann-Schwinger type. In contrast to the scattering formalism given
in the literature \cite{BCL96,GA88,GAC89,CCL03}, we do not rely on the
field modified Volkov states. Instead, we employ a more convenient set
of the field-free atomic states which is particularly advantageous for
complex atomic systems with more than one target electron.
This approach, however, can only be realized if one is able to generate a
complete set of discrete target pseudostates providing an accurate
quadrature rule. In this respect we rely on the CCC method which
demonstrated its ability to build an efficient pseudostate basis for
one and two electron targets.  This gives us confidence that we can
implement our computational scheme for nonperturbative description of
MPI in complex atomic systems such as the helium atom. Thus we should
be able to extend a very successful application of the CCC method to
the weak-filed double photoionization to the strong field domain.

We demonstrated utility of our approach for a model problem of a
square well potential and a hydrogen atom.
Also, we demonstrated that the partial ionization rates, computation
of which is usually the most difficult part of the problem, remain
stable with respect to the number of photons retained in the
calculation.

\section{Acknowledgements}

The authors acknowledge support of the Australian Research Council in
the form of  Discovery grant DP0451211.


\end{document}